\documentclass[3p,times,procedia]{elsarticle}
\usepackage{nupha_ecrc}
\usepackage{lineno}
\volume{00}

\firstpage{1}

\journalname{Nuclear Physics A}

\runauth{}

\jid{nupha}

\jnltitlelogo{Nuclear Physics A}

\usepackage{amssymb}
\usepackage[figuresright]{rotating}

\usepackage{hyperref}
\usepackage{graphicx}  % needed for figures
\usepackage{dcolumn}   % needed for some tables
\usepackage{bm}        % for math
\usepackage{amssymb}   % for math
\usepackage{amsfonts}
\usepackage{graphics}
\usepackage{grffile}   % to handle dots in graphics file names
\usepackage{epsfig}
\usepackage{units}
\usepackage[usenames]{color}
\usepackage[normalem]{ulem} % for strikethroughs (\sout{})
\usepackage{color}
\usepackage[utf8]{inputenc}
\usepackage[T1]{fontenc}
\usepackage{subfigure}

\begin{document}

\begin{frontmatter}



%% Instructions from Editor: Please use the following \dochead only in the preprint version (e-print arXiv etc.); 
%% use empty \dochead{} when submitting to Nuclear Physics A!
\dochead{XXVIIth International Conference on Ultrarelativistic Nucleus-Nucleus Collisions\\ (Quark Matter 2018)}
%% e.g. \dochead{17th International Conference on Dynamical Processes in Excited States of Solids}

\title{Pion-kaon femtoscopy in Pb$-$Pb collisions at $
\mathbf{\sqrt{{\textit s}_{\rm NN}}}=2.76$ TeV
measured with ALICE}


\author{Ashutosh Kumar Pandey for the ALICE Collaboration}

\address{Indian Institute of Technology Bombay, Mumbai, India}

\begin{abstract}
Femtoscopic correlations between charged pions and kaons for different charge combinations are measured in Pb$-$Pb collisions 
at $\mathbf{\sqrt{{\textit s}_{\rm NN}}}=2.76$ TeV with ALICE at the LHC. The three-dimensional pion-kaon ($\pi-{\rm K}$) correlation functions and double ratios
in the out-side-long pair rest frame are studied in different centrality bins. The $\pi-{\rm K}$ femtoscopic source size parameter
($R_{\rm out}$) and emission asymmetry ($\mu_{\rm out}$) are extracted. It is observed that the
average source size of the system and the emission asymmetry between pions and kaons increase from peripheral to central events.
\end{abstract}


\end{frontmatter}

%\linenumbers

\section{Non-identical particle femtoscopy}
\label{nonidfemto}Relativistic heavy-ion collisions at the Large Hadron Collider (LHC) provide an excellent
environment to study a deconfined state of quarks and gluons. Femtoscopic techniques, i.e. analyzing the momentum correlations of
produced particles at small relative momenta, are used to study the
space$-$time characteristics of the system created. Due to this
interplay and Final State Interactions (FSI) among the particles, the two-particle correlations for non-identical pairs
are sensitive to space$-$time coordinates of the particle emission points
as well as the difference in average emission points (emission asymmetry) of different particle species.

%********************************************************************************************

\section{Method}
\label{method}
The two particle correlation function is defined as
\begin{eqnarray}
 C({\bf{p}}_{\rm a},{\bf{p}}_{\rm b}) = \frac{P_2({\bf p}_{\rm a},{\bf p}_{\rm b})}{P_1({\bf p}_{\rm a})P_1({\bf p}_{\rm b})},
\end{eqnarray}
where $P_2$ is the conditional probability to observe particles with momenta ${\bf p}_{\rm a}$ and ${\bf p}_{\rm b}$ together, while $P_1$ is the probability of observing a particle with a given
momentum.
The experimental correlation function is constructed as
\begin{eqnarray}
 C({\bf k^*}) = \frac{\int N({\bf p}_{\rm a},{\bf p}_{\rm b})\delta({\bf k^*} - \frac{1}{2}({\bf p}_{\rm a}^*-{\bf p}_{\rm b}^*))d^3p_{\rm a}d^3p_{\rm b}}{\int D({\bf p}_{\rm a},{\bf p}_{\rm b})\delta({\bf k^*} - \frac{1}{2}({\bf p}_{\rm a}^*-{\bf p}_{\rm b}^*))d^3p_{\rm a}d^3p_{\rm b}}
= \frac{N({\bf k^*})}{D({\bf k^*})},
 \end{eqnarray}
where $N({\bf p}_{\rm a},{\bf p}_{\rm b})$ is the distribution when both particles originate from same event and contain correlations
while $D({\bf p}_{\rm a},{\bf p}_{\rm b})$ is the distribution when particles originate from two different events and hence contain
no correlation. The half of the relative momentum of pairs in the Pair Rest Frame (PRF) is
represented by $k^*$.

The Koonin-Pratt's equation \cite{prattkoonineqn}, which
relates the experimental correlation function with the source
emission function, is given by
\begin{eqnarray}
 C({\bf{k^*}}) = \int d{\bf{r'}} |\psi({\bf{k^*}},{\bf{r'}})| ^{2} S(\bf{r'}) .
\end{eqnarray}
The experimental correlation functions were parametrized by assuming the source function as a three dimensional Gaussian function
with three different sizes $R_{\rm out}$, $R_{\rm side}$ and $R_{\rm long}$ in Out, Side and Long directions \cite{OSL}, respectively, with the mean
value $\mu_{\rm out}$ corresponding to the emission asymmetry:

\begin{eqnarray}
 S({\bf{r}}) = exp \left( - \frac{(r_{\rm out} - \mu_{\rm out})^2}{R_{\rm out}^2} - \frac{r_{\rm side} ^2}{R_{\rm side}^2} - \frac{r_{\rm long} ^2}{R_{\rm long}^2} \right)
\end{eqnarray}
where $r_{\rm out}$, $r_{\rm side}$ and $r_{\rm long}$ are the components of the relative separation vector {\bf{r}} of the emission points.

\subsection{Double ratios}\label{sec:doubleratio}
The strength of the correlation function depends on the sign of  $\bf {k^* \cdot v}$ where
$\bf{v}$ is the pair velocity.
In the transverse plane C+ corresponds to the case where pions are faster, i.e. ${\bf {k^* \cdot v}} > 0$, and
C- corresponds to the case where kaons are faster, i.e. ${\bf {k^* \cdot v}} < 0$
  
C+ will be different from C- if both particles leave the system at different space$-$time points. For example, if kaons leave the
system earlier than pions, C+ will be stronger since faster pions will be catching up the kaons and will have larger interaction time.
In this case, C- will be weaker because both particles will be moving away and will have smaller interaction time.
The ratio C+/C- is known as double ratio and is sensitive to the emission asymmetry. If it deviates from unity,
an emission asymmetry
exists.
\subsection{Spherical harmonics decomposition of the correlation Function}\label{sec:spherical_harmonics}
	
	The Spherical Harmonics ({\bf SH}) decomposition of the correlation function is another way to study the emission
	source. In this method, the 3D correlation function is converted into an infinite set of 1D components defined as
	\begin{equation}
	  C_\mathrm{l}^\mathrm{m}(\vec{k^*}) = \frac{1}{\sqrt{4\pi}} \int d\varphi d(cos\theta) C(k^*,\theta,\varphi)Y_\mathrm{l}^\mathrm{m}(\theta,\varphi),
	\end{equation}
	where $C_\mathrm{l}^\mathrm{m}$ are the components of the correlation function and $Y_\mathrm{l}^\mathrm{m}$ are the
	spherical harmonics. Most of the real components and all imaginary components of the correlation function vanish due to the 
azimuthal symmetry \cite{adam_prc}. The $C_0^0$ and $\Re C_1^1$ component signal the source size and the average emission asymmetry, respectively.
\section{Analysis details}
The present measurements are based on the study of pion-kaon femtoscopic correlations in Pb$-$Pb collisions measured at
$\mathbf{\sqrt{{\textit s}_{\rm NN}}}=2.76$ TeV by the ALICE detector \cite{alice} in 2011. The minimum bias events were triggered by the coincidence of a signal
in the V0 detector and the Silicon Pixel Detector (SPD) of the ITS detector. The analysis has been performed for
0-5, 5-10, 10-20, 20-30, 30-40 and 40-50 \% centrality bins determined by the V0 detector. The reconstructed primary vertex along the beam direction is 
required to
lie
within $\pm$10.0 cm with respect to the center of the ALICE detector. The charged tracks were reconstructed
using the TPC detector only. The distance of closest approach (DCA)
of a track to the primary vertex in the transverse ($DCA_{xy}$) and 
longitudinal ($DCA_{z}$) directions are required to be less than 2.4 cm and
3.2 cm, respectively, to reduce the
contamination from  secondary tracks emanating from weak decays and
from interactions with the detector material. Tracks with a transverse momentum within 0.19 GeV/$c$ <
$p_{\rm T}$ < 1.5 GeV/$c$ measured in the pseudo-rapidity range $|\eta| < 0.8$ are
selected. Combined information from TPC and TOF is used to
identify charged tracks as pions and kaons.
All track pairs, which share more than 5\%
of their total hits in the TPC, are excluded from the analysis.  Another selection on the fraction of merged points was
used to find the fraction of two tracks ($|\Delta\eta| < 0.1$), which
were closer than the average cluster size (3 cm) in the TPC. The merged 
fraction is defined as the ratio of the number of points for which the 
distance between the tracks is less than 3 cm to the total number of 
steps of 3 cm calculated in the considered TPC radius range. 
The unlike-sign (like-sign) pairs with a merged fraction above 1\% (
7\%) were removed.
To remove contamination of
$e^{-}-e^{+}$ pairs originating from $\gamma$ conversions, pairs are removed 
if the invariant mass of pairs with assumed electron mass and the
difference of polar angle $\delta\theta$ between the tracks was less
than 0.002 GeV/$c$ and  0.008 radians, respectively.
The uncorrelated pair background is constructed by pairing tracks from 
different events. 

\section{Results}
%\begin{figure}
%\centering
%\includegraphics[scale=0.14]{mult_rhic_auau.eps} 
%\includegraphics[scale=0.20]{2018-May-11-CorrFcn_CC.eps}
%\includegraphics[scale=0.16]{2018-May-11-SHCorrFun.eps}
%\caption{Left and middle figure: The pion-kaon correlation function for Pb$-$Pb 
  %collisions at $\sqrt{s_{NN}}$ = 2.76 TeV}
%\label{fig1}
%\end{figure}

\begin{figure}
\centering
\includegraphics[scale=0.35]{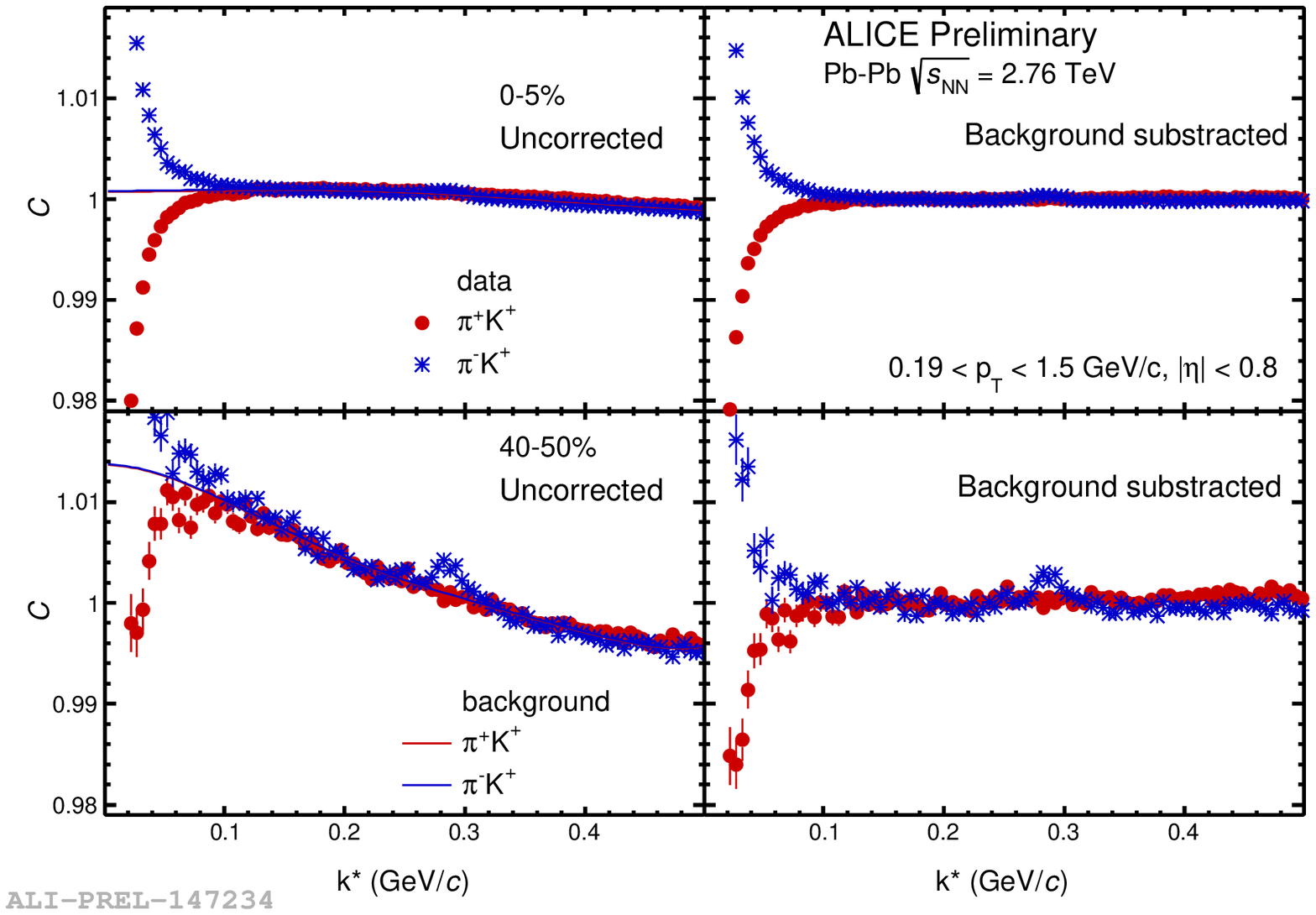}
\includegraphics[scale=0.35]{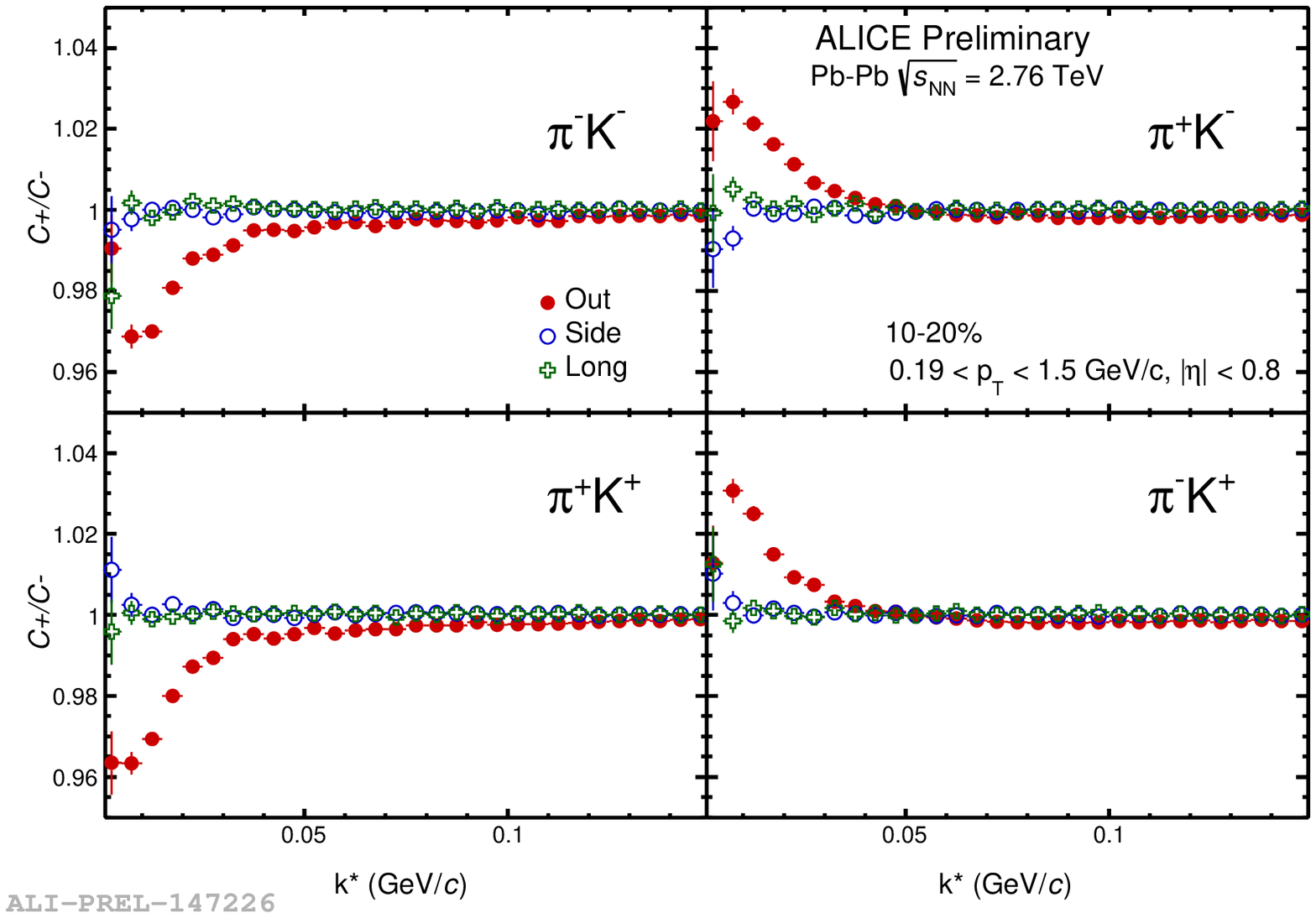}
\caption{Left: The pion-kaon correlation function to show
the effect of non-femtoscopic background. The background fit corresponds to sixth order
polynomial function. Right: The pion-kaon double ratio for Pb$-$Pb 
  collisions at $\mathbf{\sqrt{{\textit s}_{\rm NN}}}=2.76$ TeV. (Errors are statisitical only)}
\label{fig2}
\end{figure}

\begin{figure}
\centering
\includegraphics[scale=0.35]{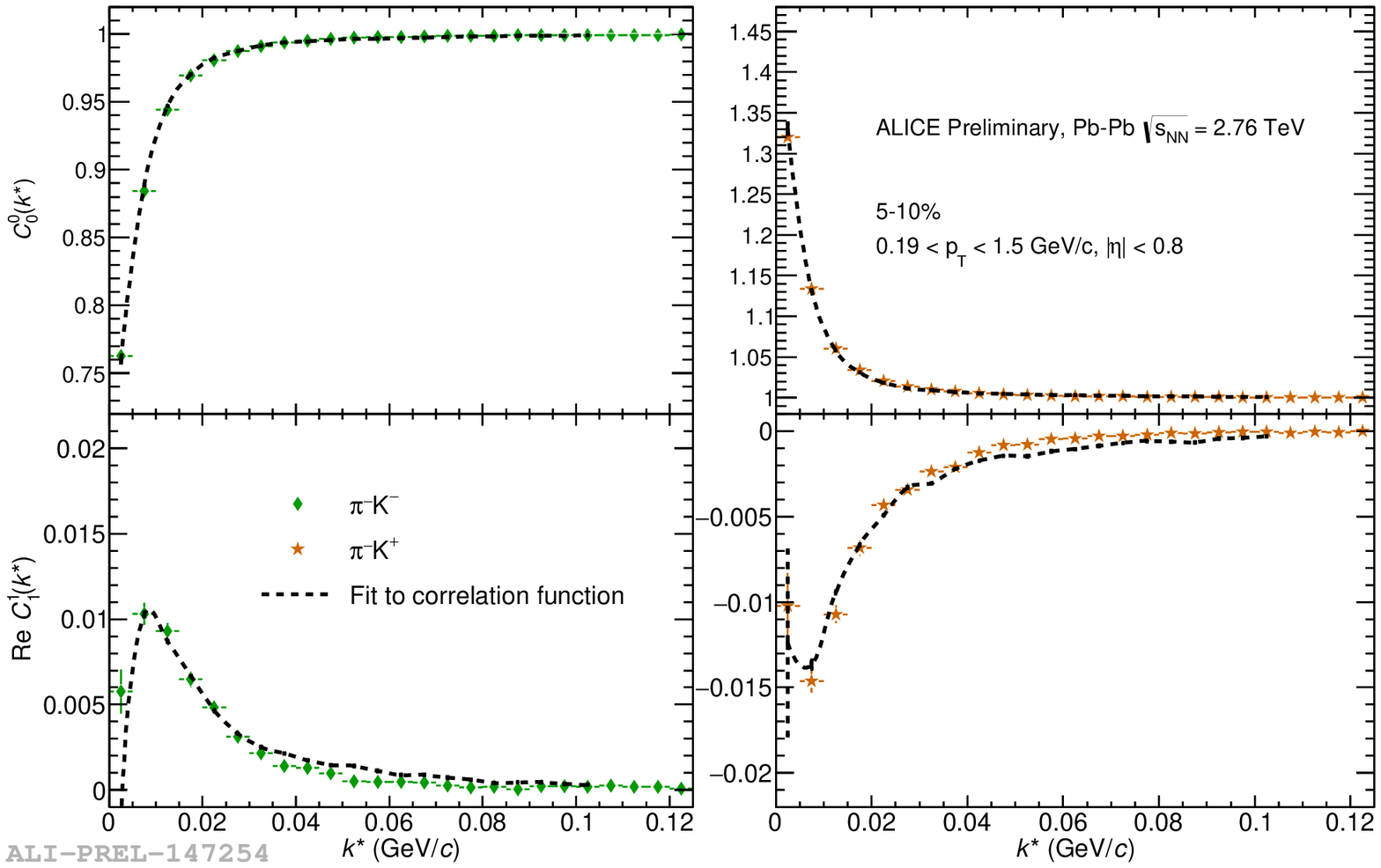}
\includegraphics[scale=0.35]{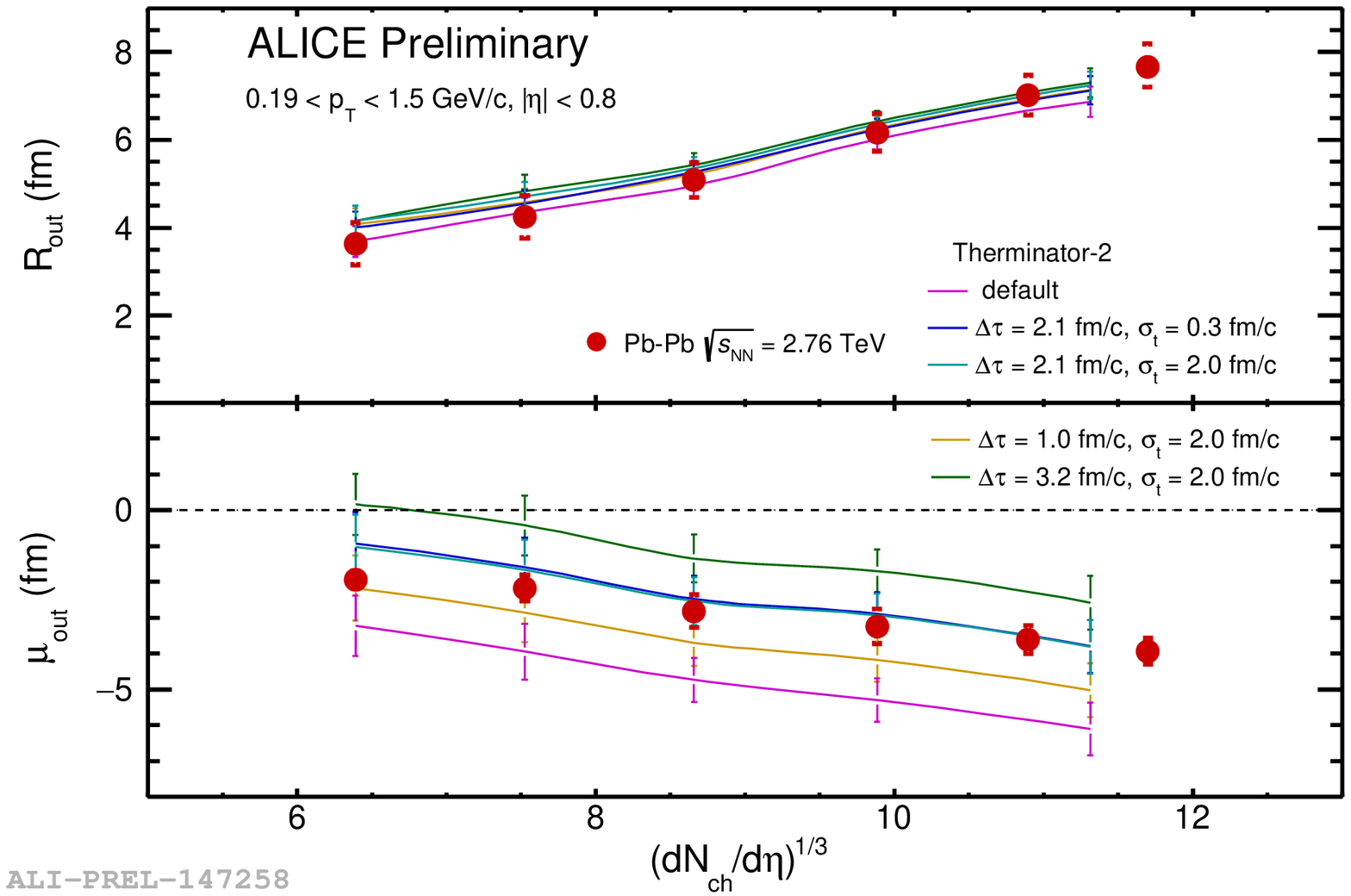}
\caption{Left: The pion-kaon correlation function for all charge combinations with their fits computed
using CorrFit software (Errors are statisitical only); Right: Source size (upper panel) and pion-
kaon emission asymmetry (lower panel) from pion-kaon correlation functions for Pb$-$Pb collisions at $\mathbf{\sqrt{{\textit s}_{\rm NN}}}=2.76$ TeV
as a function of $(dN_{ch}/d\eta)^{1/3}$ .
The solid lines show predictions from calculation of source-size and
emission asymmetry using Therminator-2 model with default and selected
values of additional time delays for kaons (Errors are statisitical and systematical).}
\label{fig3}
\end{figure}

The left panels of Fig. \ref{fig2} and \ref{fig3} show the correlation functions, in the Cartesian coordinate and Spherical Harmonics representation
for all possible charge combinations of pion-kaon
pairs, respectively. 
The strength of C+, C- and
$C^{0}_{0}$ is above
unity for unlike-sign pairs and below unity for like-sign pairs suggesting an
attractive and repulsive Coulomb interaction between the pairs,
respectively. The non-vanishing signal values of the $\Re C^{1}_{1}$
component and the deviation of C+/C- from unity indicates the presence of an emission asymmetry. In other words, on average
pions and kaons
are not emitted at the same source radius or time. 

It can be observed from Fig. \ref{fig2} that the correlation function is 
not exactly flat at unity in the non-femtoscopic region. This is 
essentially due to the presence of other event wide pair correlations 
like
elliptic flow $v_2$, global event energy and momentum conservation, resonance decay correlations, residual correlation,
jets etc. Therefore, 
one needs to isolate these background correlations before extracting 
the source parameters. 
The procedure to estimate the non-femtoscopic background is described 
in \cite{adam_bkg}, where it is shown that the behavior of the 
non-femtoscopic baseline can be characterized by a 
sixth order polynomial function. The same procedure was used to correct the 
effect of non-femtoscopic background in the present 
analysis. 

A numerical procedure as employed in the STAR experiment was  followed 
to extract source size and emission asymmetry \cite{adam_prc, corrfit} by assuming the source function as in Eq. (4).
Using the mentioned source function, one can numerically integrate the 
Eq. (3) with the corresponding wave function (indicating the
type of interactions) to calculate the correlation function.
The input momentum distributions are taken from real pairs to ensure
that the momentum acceptance is same. The calculated correlation is compared to
the measured via a $\chi^{2}$ test. The procedure is repeated for
different sets of source parameters to arrive at the best agreement
with the measured data. 
In this work, the CorrFit software package \cite{corrfit} was used to perform the 
numerical fitting described above and extract the experimental
${R_{\rm out}}$ and $\mu_{\rm out}$.

The right plot in Fig. \ref{fig3} shows the pion-kaon source size and emission
asymmetry as a function of the cube root of charged particle multiplicity 
density  in Pb$-$Pb collisions at $\mathbf{\sqrt{{\textit s}_{\rm NN}}}=2.76$ TeV. One observes that the system size increases from peripheral to central events. The extracted emission asymmetry is also observed to increase
with particle multiplicity and is negative. This implies that pions
are emitted closer to the centre of the source.
The results are compared to the predictions from the
Therminator2 model \cite{therm2} with the default calculations and calculations
with selected values of time delay for kaon emission \cite{adam_noniden}. 
This was motivated by the  recent measurements of identical kaon
femtoscopy which showed that kaons are emitted, on average, 2.1 fm/$c$ 
later than pions. The default Therminator2 calculations overpredict
the experimental data while an introduction of a time delay of 2.1 fm/$c$
in kaon emission decreases the asymmetry and is in good agreement with
the experimental measurement.
\section{Conclusion}
The first measurements of pion-kaon femtoscopic correlations in Pb$-$Pb 
collisions at $\mathbf{\sqrt{{\textit s}_{\rm NN}}}=2.76$ TeV have been performed. The radius of the source ${R_{\rm out}}$ shows a decreasing trend from central to peripheral
collisions. A finite emission asymmetry is also observed 
which shows a similar trend as the radius. The results are consistent with the Therminator2 coupled with (3+1)D viscous hydrodynamic
model calculations of pion-kaon 
emission asymmetry when an additional time delay of 2.1 fm/$c$ is
introduced for the kaons.



%% References


%\bibliographystyle{elsarticle-num}

\begin{thebibliography}{50}
\medskip
\bibitem{prattkoonineqn} S.E. Koonin, PLB70 (1977) {\bf{43}}, S. Pratt {\it et al.}, PRC42 (1990) 2646
\bibitem{OSL}Michael Annan Lisa,  Scott Pratt,  Ron Soltz,  Urs Wiedemann, {\bf  Annual Review of Nuclear and Particle Science }, Vol. 55:357-402 (2005)
\bibitem{adam_prc}A.~Kisiel, Phys. Rev. {\bf C 81}, 064906 (2010).
\bibitem{alice} K. Aamodt {\it et al}. The ALICE experiment at the CERN LHC. JINST, 3:S08002, 2008.
\bibitem{adam_bkg} A. Kisiel, Acta.Physica.Polonica {\bf B 48}, (2017). 
\bibitem{corrfit}A. ~Kisiel, NUKLEONIKA, {\bf 49} , S81-S83, (2004)
\bibitem{therm2} M. ~Chojnacki, A.~Kisiel, W.~Florkowski and W. ~Broniowski, Comput. Phys. Commun. {\bf 183},746 (2012).  
\bibitem{adam_noniden} A. Kisiel, {\bf arXiv:1804.06781}, (2018). 
\end{thebibliography}






% \end{thebibliography}

\end{document}